# Performance Analysis and Optimization for Laser-Phase-Noise based Quantum Random Number Generation

Jinlu Liu[1,2], Jie Yang[2,*], Yu Gao[2], Guowei Zhang[1], Yan Pan[2], Heng Wang[2], Yuyang Ding[3], Yang Li[2], Wei Huang[2], Bingjie Xu[2,†], Wei Chen[1,4,‡]

[1]CAS Key Laboratory of Quantum Information, University of Science and Technology of China, Hefei 230026, China.
[2]National Key Laboratory of Security Communication, Institute of Southwestern Communication, Chengdu 610041, China.
[3]Hefei Guizhen Chip Technologies Co. Ltd., Hefei 230000, China.
[4]Hefei National Laboratory, University of Science and Technology of China, Hefei 230088, China.
[*] everest1573@163.com
[†] xbjpku@163.com
[‡] weich@ustc.edu.cn

## Abstract

The quantum random number generation based on laser phase noise, which is featured with high random number generation rate and ease for photonic integration, has been extensively investigated and demonstrated. Despite these advancements, a theoretical model to achieve optimal performance in terms of maximizing the random number generation rate is still incomplete. In this work, a comprehensive physical model for this scheme is introduced to accurately predict the power spectrum of entropy source and probability distribution of raw data, based on which the entropy source bandwidth and extractable randomness can be accordingly estimated and thus the system performance can be quantitatively evaluated and optimized. The model is sufficiently validated through both simulation and experiment with significant agreement under various typical setups. Furthermore, our proposal enables the proactive design of experimental parameters to achieve optimal system performance, which is crucial for the design and practical implementation of photonics integrated quantum random number generations.

## I. Introduction

Random numbers serve as a foundational resource for modern science and technology, particularly in fields of statistical simulations and secure communications [1~3]. As a promising solution for generating true random numbers, the quantum random number generation (QRNG), which exploits intrinsic randomness of quantum processes, has been demonstrated of significant advantages in producing high-quality random numbers with theoretically provable randomness. Various QRNG schemes utilizing different physical entropy sources have been demonstrated, such as photon arrival times [4~6], quantum non-locality [7], laser phase noise [8~26], vacuum noise [27~34] and amplified spontaneous emission noise [35~37]. Particularly, the laser-phase-noise (LPN) based QRNGs have attracted considerable interest owing to the high random number generation rate [8, 14, 25] and ease of photonic integration [15, 19~22, 25]. Currently, the LPN-based QRNGs implemented through off-the-shelf components and photonic integration circuits have been demonstrated with random number generation rate over 100Gbps [25] and 10Gbps [22], respectively. Fundamentally, the random number generation rate depends on both the extractable randomness and the sampling rate, with the latter determined by the entropy source bandwidth. While the extractable randomness has been widely studied [5, 10, 18, 38], the investigation into the entropy source bandwidth, specifically regarding the frequency-domain characteristics of the entropy source, remains lacking. Hence, a comprehensive theoretical model for system performance optimization in terms of maximizing the random number generation rate remains incomplete.

In this work, a systematic physical model for LPN-based QRNGs, incorporating crucial parameters including the linewidth of the laser $\Delta v$, time delay difference of unbalanced Mach-Zehnder interferometer (uMZI) $\tau_l$, and digitization resolution $n$ and dynamic range $R$ of analog-to-digital converter (ADC), is proposed. Compared with existing works, the proposed approach enables the simultaneous characterization of both entropy source bandwidth and extractable randomness, providing a theoretical framework for the proactive optimization of the random number generation rate. This methodology, rigorously validated through theoretical simulations and experiments, is particularly valuable for the development of photonic integrated QRNGs. By enabling precise performance prediction before hardware realization, this work offers a strategic path to avoid the expensive and iterative tape-out cycles typically associated with chip-scale designs.

# II. Model

For a general QRNG system, the system performance can be estimated by [25, 31]

$$K = f_s \cdot H_{min} \leq 2 \cdot B_{ES} \cdot H_{min}, \qquad (1)$$

where $K$ is the random number generation rate, $f_s$ is the sampling rate applied to the entropy source signal, which is generally bounded by the entropy source bandwidth $B_{ES}$ with $f_s \leq 2B_{ES}$, and $H_{min}$ is the min-entropy of quantum noise per sample. Here the assessment of $H_{min}$ assumes that the sampled data are independent and identically distributed (i.i.d.). Based on Eq.(1), to achieve the optimal system performance, one needs to maximize the product of $B_{ES}$ and $H_{min}$, while simply maximizing either parameter individually is sub-optimal.

Typically, the $B_{ES}$ is determined by identifying the 3-dB cutoff point of the power spectrum, while $H_{min}$ is evaluated via $H_{min} = -\log_2(P_{max})$. $P_{max}$ corresponds to the maximum probability that an eavesdropper could correctly predict the measurement outcomes. Practically, $P_{max}$ can be estimated from the discrete probability distribution obtained by quantizing the entropy source signal using a general analog-to-digital convertor (ADC) [10, 38]. Accordingly, establishing a comprehensive model capable of characterizing quantum noise in both the time and frequency domains is fundamental to a *priori* performance evaluation of QRNG systems. Such a model serves as the essential framework for optimizing system performance under various experimental configurations.

In this work, a comprehensive model that explicitly incorporates critical system parameters to illustrate both the power spectrum of entropy source and the probability distribution of raw data is established, which enables the simultaneous estimation of $B_{ES}$ and $H_{min}$ that depend directly on system parameters, such as $\Delta\nu$ and $\tau_l$. Based on the proposed model, one can identify the experimental configuration required to achieve optimal system performance.

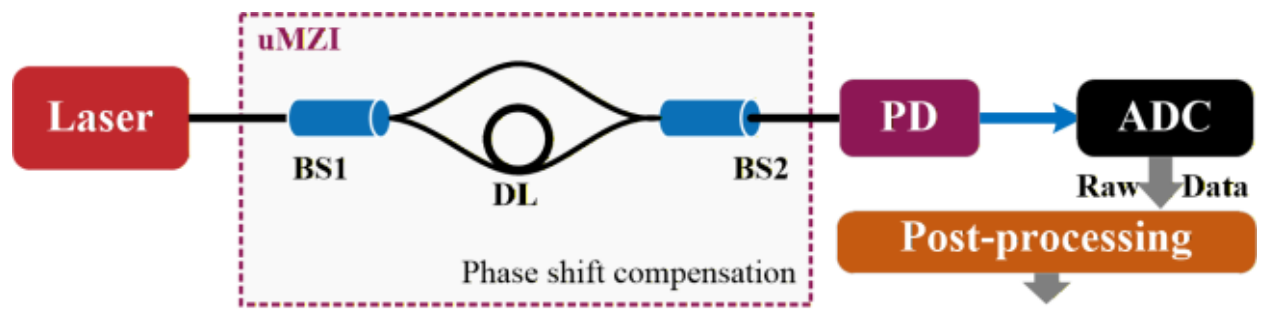

**Figure 1 Typical setup of the LPN-based QRNG**

As shown in Figure 1, a typical setup of the LPN-based QRNG includes a laser, an uMZI, a photo-detector (PD), an ADC and a post-processing module. The random phase difference of optical signal emitted from the laser at two discrete times $\Delta\theta$, is converted into optical intensity fluctuations via the uMZI，which are subsequently converted to electronic signals by the PD. By employing an active feedback control, the random phase shift introduced by uMZI can be compensated. Then the DC-filtered output signals of PD, which is proportional to $\sin\Delta\theta$ termed as 'quantum noise', can be expressed by

$$Q(t, \tau_l) = A\sin\Delta\theta(t, \tau_l), \qquad (2)$$

where $A$ is a conversion factor totally accounts for the amplitude of the optical signal, the attenuation introduced by the uMZI and the responsibility of the PD, $\tau_l$ is the time delay difference of the uMZI and $\Delta\theta$ is the 'phase noise', i.e. $\Delta\theta(t, \tau_l) = \theta(t + \tau_l) - \theta(t)$.

In previous works [13, 14, 20], $\Delta\theta(t, \tau_l)$ is directly modeled as a Gaussian random variable with variance $\sigma^2_{\Delta\theta,\tau_l} = 2\pi \cdot \Delta\nu \cdot \tau_l$, where $\Delta\nu$ is the laser linewidth. This model establishes the relationship between the phase noise variance and experimental parameters. Nevertheless, it fails to elucidate how the phase noise evolves under various experimental parameters and subsequently affects the system performance. In this work, we employ the Wiener process to model laser phase, thereby characterizing the evolution of the system output signal in both the time and frequency domain [26]. Since the laser phase $\theta(t)$ physically executes Brownian motion, the variance of the phase noise over an interval of $t$ can be linearly approximated by [39]

$$\sigma_{\Delta\theta}^2 = \frac{R_{sp}}{2s}(1+\alpha^2)t, \quad (3)$$

where $R_{sp}$ is the spontaneous emission rate, $s$ is the average number of photons in the cavity, and $\alpha$ is the linewidth enhancement factor. Mathematically, the phase $\theta(t)$ is a Wiener process. Based on the definition of the Wiener process, the instantaneous phase $\theta(t)$ of an optical signal with a Lorentzain spectrum can be represented in discrete-time as $\theta(m\tau_s)$, which satisfies

$$\theta\big((m+1)\tau_s\big) - \theta(m\tau_s) \sim N(0, 2\pi\Delta v\tau_s), \quad (4)$$

where $\tau_s$ denotes the sampling period, $\theta\big((m+1)\tau_s\big)$ is an essential delay term of $\theta(m\tau_s)$ with $m = 0,1,2,3\cdots$. Then the discretized quantum noise can be expressed as

$$Q(m\tau_s, \tau_l) = A\sin\{\theta(m\tau_s + k\tau_s) - \theta(m\tau_s)\}, (5)$$

where $k\tau_s = \tau_l$ and $k$ is a constant.

Based on the theoretical model established in Eqs.(2 ~5), it is achievable to numerically construct the intrinsic quantum noise, and subsequently simulate the entropy source signal and the digitized raw data. Furthermore, this numerical framework allows us to systematically evaluate the system performance and determine the optimal physical parameters for maximizing the random number generation rate. The detailed simulation procedure is provided in Supplement A.

# III. Results

To bridge the theoretical framework with practical implementations, we conduct a systematic experimental evaluation utilizing the experimental setup depicted in Fig. 2. A coherent, continuous-wave beam emits from a laser diode, driven by a commercial laser driver. The laser's linewidth $\Delta v$ can be tuned via precise adjustment of the driving current. Subsequently, the beam is coupled into the uMZI after adjusted by a variable optical attenuator (VOA). Inside the interferometer, the time delay difference $\tau_l$ is adjusted via varying length of delay fiber, while a phase modulator and an additional VOA are employed to compensate for phase shifts and balance the optical power between two paths, respectively. Then, the interference results output from uMZI are split into two parts, with one portion detected by a 1.6 GHz photo-detector (PD1) and then sampled to generate raw data by an 8-bit ADC, which is realized by an oscilloscope (KeySight, DSOV084A, 8 GHz bandwidth) to obtain raw data, while another portion is detected by PD2 to provide a feedback signal for stabilizing the phase shift of the uMZI.

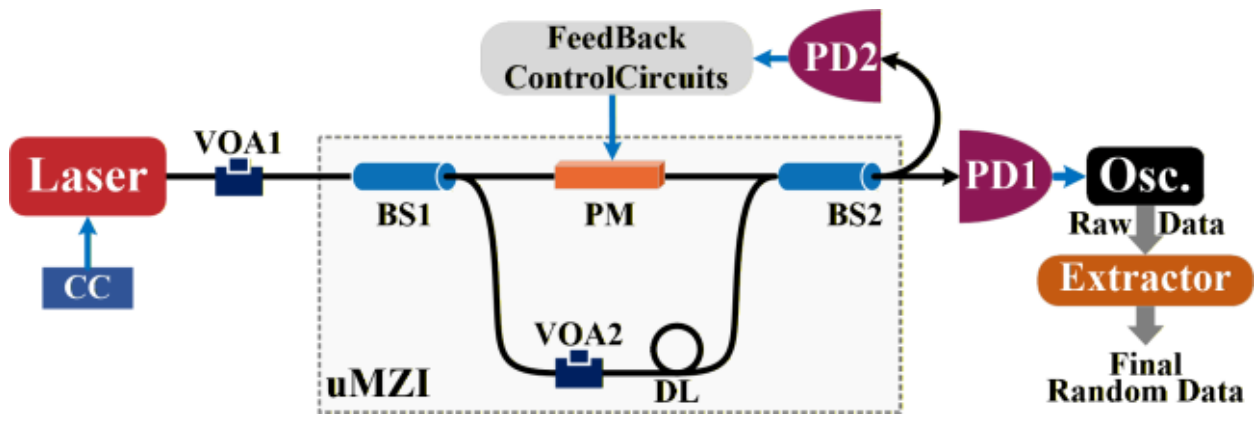

**Figure 2 The experimental setup of LPN-based QRNG**

**CC: current controller, BS: beam splitter, PM: phase modulator, PD: photo-detector, uMZI: unbalanced Mach-Zehnder interferometer, DF: delay fiber, VOA: variable optical attenuator, ADC: analog-to-digital convertor**

Based on the proposed model, $\Delta v$ and $\tau_l$ are the dominant parameters for quantum noise $Q$, the experiments are configured into two scenarios: (a) fixed $\Delta v$ with different $\tau_l$ and (b) fixed $\tau_l$ with different $\Delta v$.

Utilizing the measurements acquired from the aforementioned setup, we systematically analyze the system performance. The accuracy of the model is firstly verified through a direct comparison between the theoretical and measured entropy source power spectra and raw data distributions. Subsequently, building upon this validated model, the reliable extraction of key parameters, specifically the $B_{ES}$ and $H_{min}$, is demonstrated. Ultimately, this quantitative evaluation framework is utilized to actively optimize the experimental configuration, achieving the maximum random number generation rate.

## A. Model validation

The theoretical model is validated through a comprehensive assessment of the entropy source across the frequency and time domains. Starting with the frequency-domain analysis, Fig.3 presents a comparison between the simulated and experimental power spectra of the entropy source for scenarios (a) and (b), exhibiting excellent agreement. The successful validation provides a reliable foundation for subsequent estimation of the $B_{ES}$.

Notably, in this work, the LPN-based QRNG is analyzed within the framework of device-dependent implementations. To estimate the entropy source bandwidth, it is necessary to calibrate the classical noise in frequency domain first. The detailed results are shown in Supplement D, which demonstrated that the electronic noise of the photodetector is overwhelming in the classical noise and the measurement total noise is dominated by the quantum noise. Furthermore, the bandwidth of raw data is determined by the component with the narrowest frequency response among the entropy source (including photo-detector) and ADC. Given that both the detector and oscilloscope exhibit flat responses beyond 1GHz, the spectral roll-off of raw data depicted in Fig.3 is intrinsically attributed to the entropy source itself.

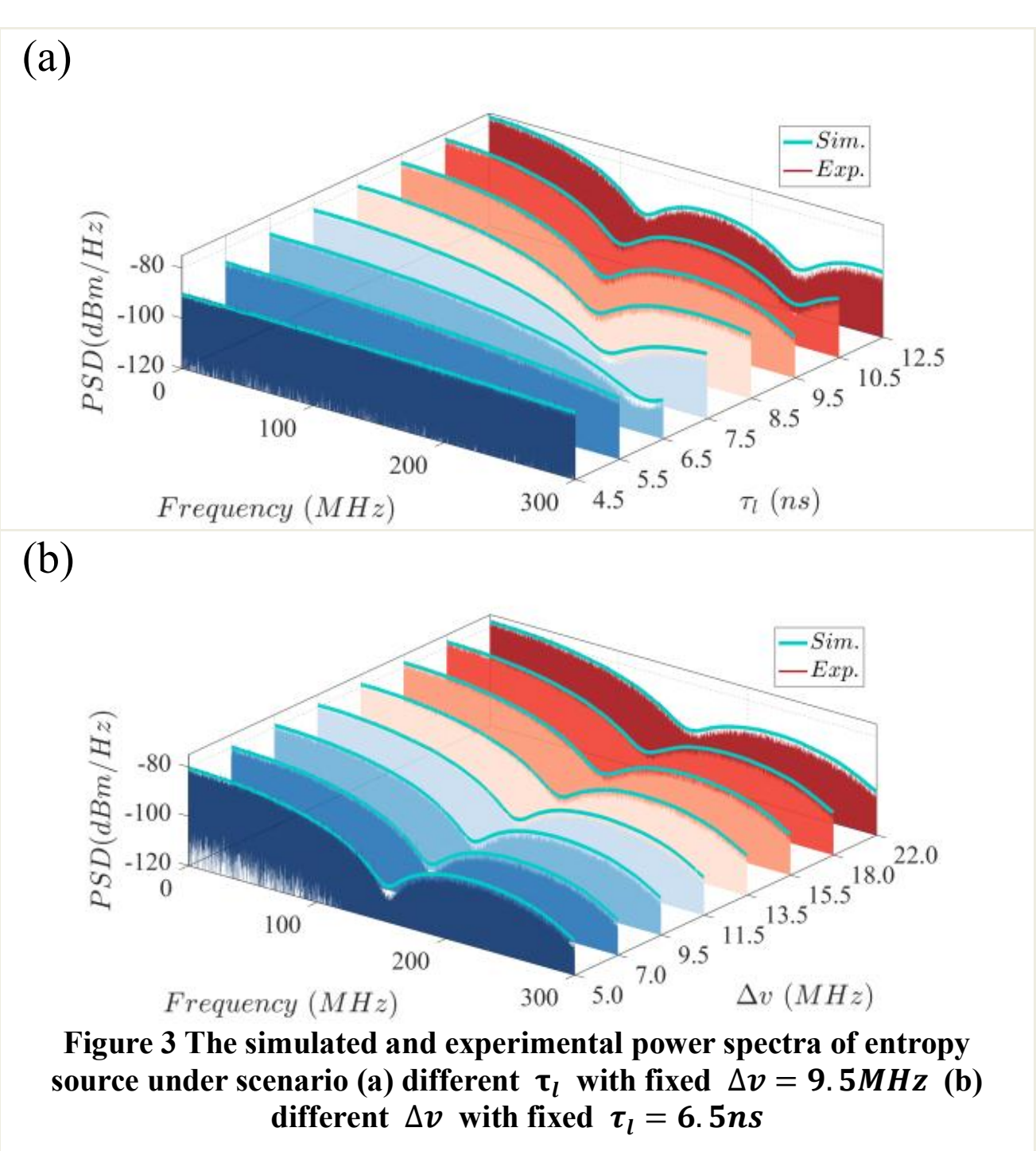

**Figure 3 The simulated and experimental power spectra of entropy source under scenario (a) different $\tau_l$ with fixed $\Delta v = 9.5MHz$ (b) different $\Delta v$ with fixed $\tau_l = 6.5ns$**

A qualitative observation of Fig.3 (a) reveals that the $B_{ES}$ gradually decreases with an increasing $\tau_l$. In contrast, Fig.3 (b) demonstrates that $B_{ES}$ exhibits no significant variation as $\Delta v$ increases. A detailed quantitative analysis of the impacts induced by $\tau_l$ and $\Delta v$ on $B_{ES}$ is deferred to the subsequent subsection. Presently, we proceed to complete the model validation in the time domain.

In the time domain, the theoretical model is validated through a rigorous assessment of the statistical characteristics of the raw data. Fig.4 compares the simulated and experimental probability distributions of the raw data for both scenarios.

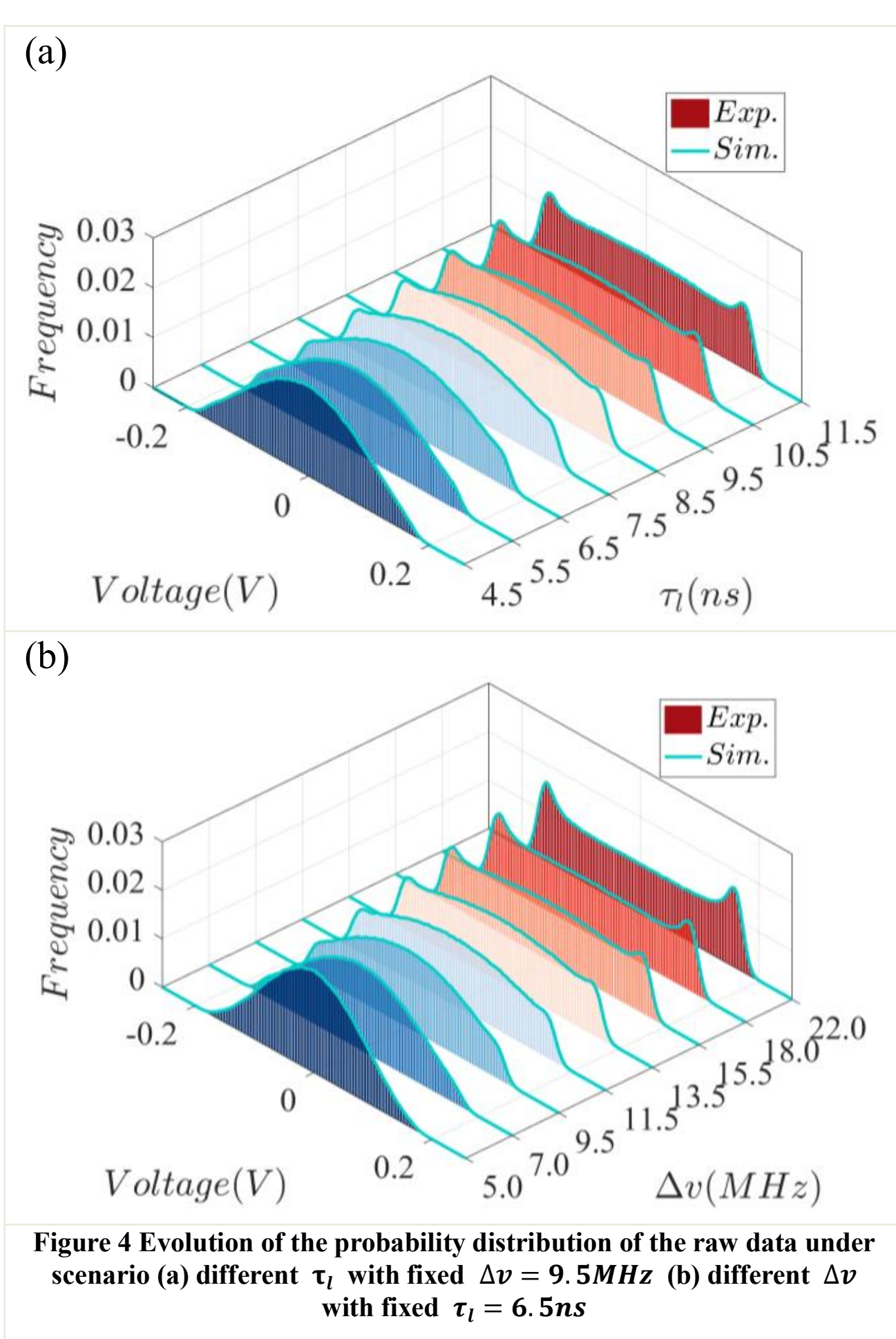

**Figure 4 Evolution of the probability distribution of the raw data under scenario (a) different $\tau_l$ with fixed $\Delta v = 9.5MHz$ (b) different $\Delta v$ with fixed $\tau_l = 6.5ns$**

Specifically, under scenario (a), as shown in Fig.4(a), the probability distribution is consistently symmetry about the zero voltage across the variation range of $\tau_l$. Initially, when $\tau_l$ is relatively small, which correspondingly results in a small variance of phase noise $\sigma^2_{\Delta\theta,\tau_l}$, the probability distribution exhibits quasi-Gaussian, indicating that the majority of the raw data concentrates around the mean. However, as $\tau_l$ increases, which results in a larger $\sigma^2_{\Delta\theta,\tau_l}$ and $\Delta\theta$ gradually exceeds outside $[-\pi, \pi]$, the distribution progressively accumulates toward the boundaries while decreasing at the mean. Finally, it exhibits a bimodal shape [40]. The similar results of probability distribution are achieved under scenario (b), as shown in Fig.4 (b).

The excellent agreement between the simulations and experiments under both scenarios not only validates the model's ability to predict the probability distribution of the raw data, but also ensures its reliability for accurately evaluating the $H_{min}$ in the subsequent analysis.

## B. Characterization of Bandwidth and Min-entropy

To fully characterize how the system performance scales with the parameters, based on the validated model, quantitative analysis of the impacts on $B_{ES}$ and $H_{min}$ induced by $\tau_l$ and $\Delta v$ are performed. The evolution of $B_{ES}$ and $H_{min}$ for both scenarios are presented in Fig.5 and Fig.6, respectively.

As illustrated by the cyan spheres in Fig.5, the $B_{ES}$ decreases rapidly as $\tau_l$ increases under scenario (a). This result stems from the fact that the power spectrum of the entropy source is primarily determined by $\theta\left((m+1)\tau_s\right) - \theta(m\tau_s)$, as defined by Eq.(4). For a constant $\Delta v$, a shorter $\tau_l$ (corresponding to a smaller $k$) effectively captures the phase noise over a shorter time interval. This preserves more high-frequency components in the resulting quantum noise, thereby broadening $B_{ES}$. Regarding scenario (b), where $\tau_l$ is kept constant (indicated by the pink spheres in Fig. 5), the $B_{ES}$ exhibits a slightly increase, rising from 65.2MHz to 67.2MHz, as $\Delta v$ is tuned from 5 to 22MHz. This specific tuning range is practically achievable by adjusting the driving current of a stable continuous-wave laser. Fundamentally, a broaden $\Delta v$ accelerates phase diffusion process, which introduces more high-frequency spectral components into the measured signal and thus broadening $B_{ES}$.

Notably, while driving the laser below its threshold could further increase $\Delta v$, it would significantly degrade the quantum-to-classical noise ratio. Furthermore, since the proposed model assumes a Lorentzian lineshape, the estimation of $B_{ES}$ remains valid only as long as the laser output adheres to this characteristic.

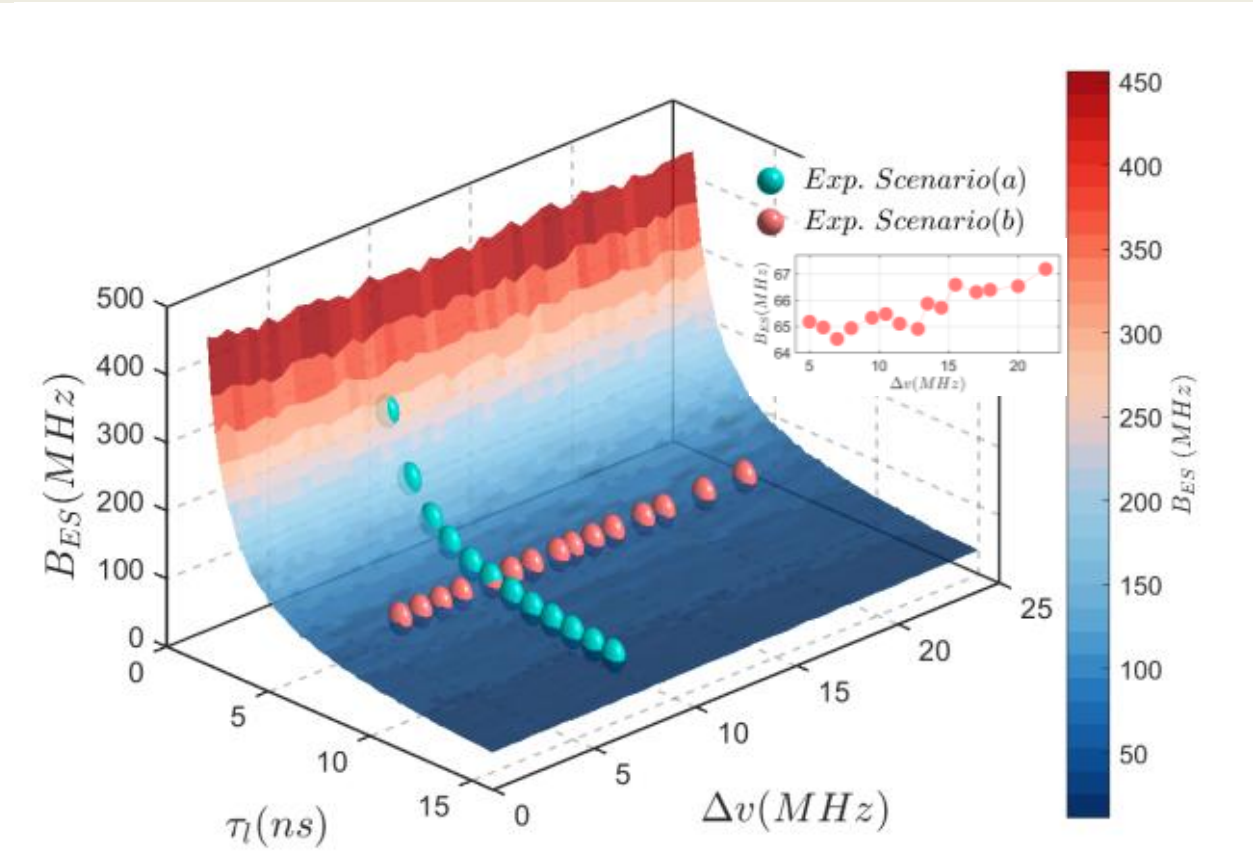

**Figure 5 Dependence of $B_{ES}$ on $\tau_l$ and $\Delta v$. The color surface represents the simulated results across the entire parameter space, where the color gradient indicates the magnitude of $B_{ES}$. Experimental data for Scenario (a) and Scenario (b) are denoted by cyan and pink spheres, respectively.**

As illustrated in Fig.5, the predicted $B_{ES}$, which is obtained by simulation and represented by the continuous color surface, are in excellent agreement with the experimental results. This consistency establishes the proposed model as an efficient approach for designing LPN-based QRNG with targeted $B_{ES}$, and a reliable guide for configuring the sampling rate.

To comprehensively evaluate the system performance, it is imperative to estimate the $H_{min}$ simultaneously. Because the raw data inevitably contaminated by classical noise, it is difficult to directly measure the pure quantum noise. However, the proposed model validated in the preceding subsections allows one to decouple and accurately characterize the quantum noise. Employing an $n$-bit ADC with quantification range of $[-V_R - \delta/2, V_R - \delta/2]$, with a bin width $\delta = V_R/2^{n-1}$, the probability distribution of quantum noise can be evaluated. Similar to the raw data, as demonstrated in Fig.4, its distribution transitions prominently from a quasi-Gaussian to a bimodal profile. Utilizing the dynamic variations in the probability distribution, the $H_{min}$ can be quantitatively assessed. The detailed methodology is provided in the Supplement B and Supplement C. The corresponding results are illustrated in Fig.6.

As illustrates by the cyan and pink spheres in Fig.6, in both scenarios for experiment, as the $\Delta v$ (or $\tau_l$) increases, the $H_{min}$ gradually increases, then reaches its maximum, and subsequently declines. This result arises from the fact that $P_{max}$ varies non-monotonically as the $\Delta v$ (or $\tau_l$) increases. The predicted $H_{min}$, which is obtained by simulation and represented by the continuous color surface, are in excellent agreement with the experimental results. This consistency establishes the proposed model as an efficient approach for designing LPN-based QRNG with targeted $H_{min}$.

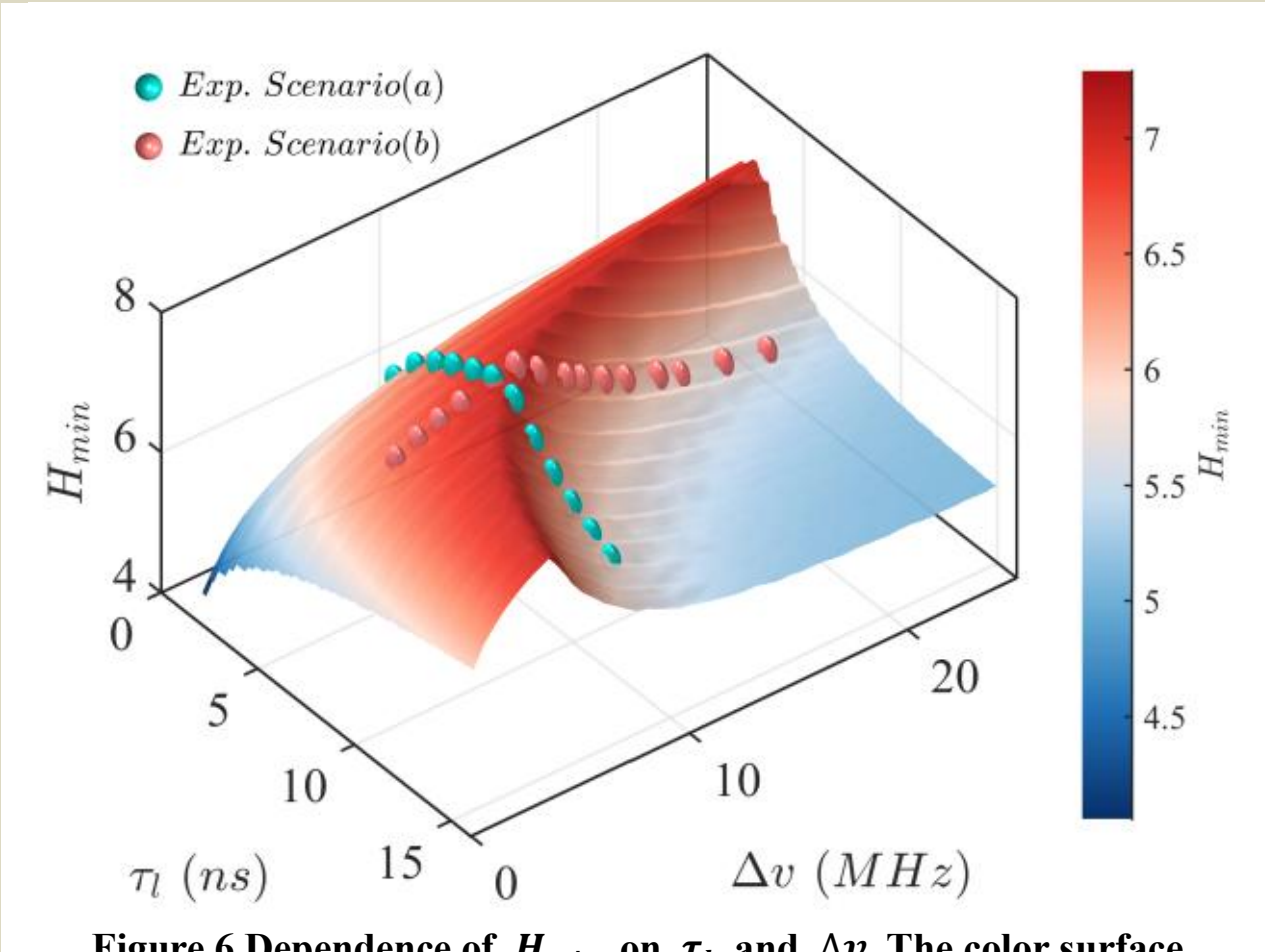

**Figure 6 Dependence of $H_{min}$ on $\tau_l$ and $\Delta v$. The color surface represents the simulated results across the entire parameter space, where the color gradient indicates the magnitude of $H_{min}$. Experimental data for Scenario (a) and (b) are denoted by cyan and pink spheres, respectively.**

Crucially, a comprehensive analysis of both scenarios reveals that $H_{min}$ reaches its peak whenever the product of $\Delta v$ and $\tau_l$ converges to a specific value. Fundamentally, this behavior stems from the fact that $H_{min}$ is inherently governed by the underlying Gaussian phase fluctuation, despite being remapped as quantum noise. Consequently, the variance $\sigma^2_{\Delta\theta,\tau_l} = 2 \cdot \pi \cdot \Delta v \cdot \tau_l$ serves as the intrinsic determinant of $H_{min}$. Under a given ADC quantization scheme, the maximum $H_{min}$ always corresponds to specific variance value. Physically, the maximization of $H_{min}$ signifies an optimal state of coherence decay. Since spontaneous emission drives the random diffusion of the optical phase, the $\sigma^2_{\Delta\theta,\tau_l}$ directly and precisely quantifies the total amount uncertainty accumulated during the detection interval.

### C. Performance optimization

While the individual optimization of either $B_{ES}$ and $H_{min}$ can be achieved based on the proposed model, maximizing the final random number generation rate $K$ requires a delicate balance between $B_{ES}$ and $H_{min}$. Across the range of attainable experimental parameters, the corresponding $K$ are determined by systematically sweeping the parameter space of $\tau_l$ and $\Delta v$. The simulated and predicted results, alongside the experimental results for both scenarios, are presented in Fig.7.

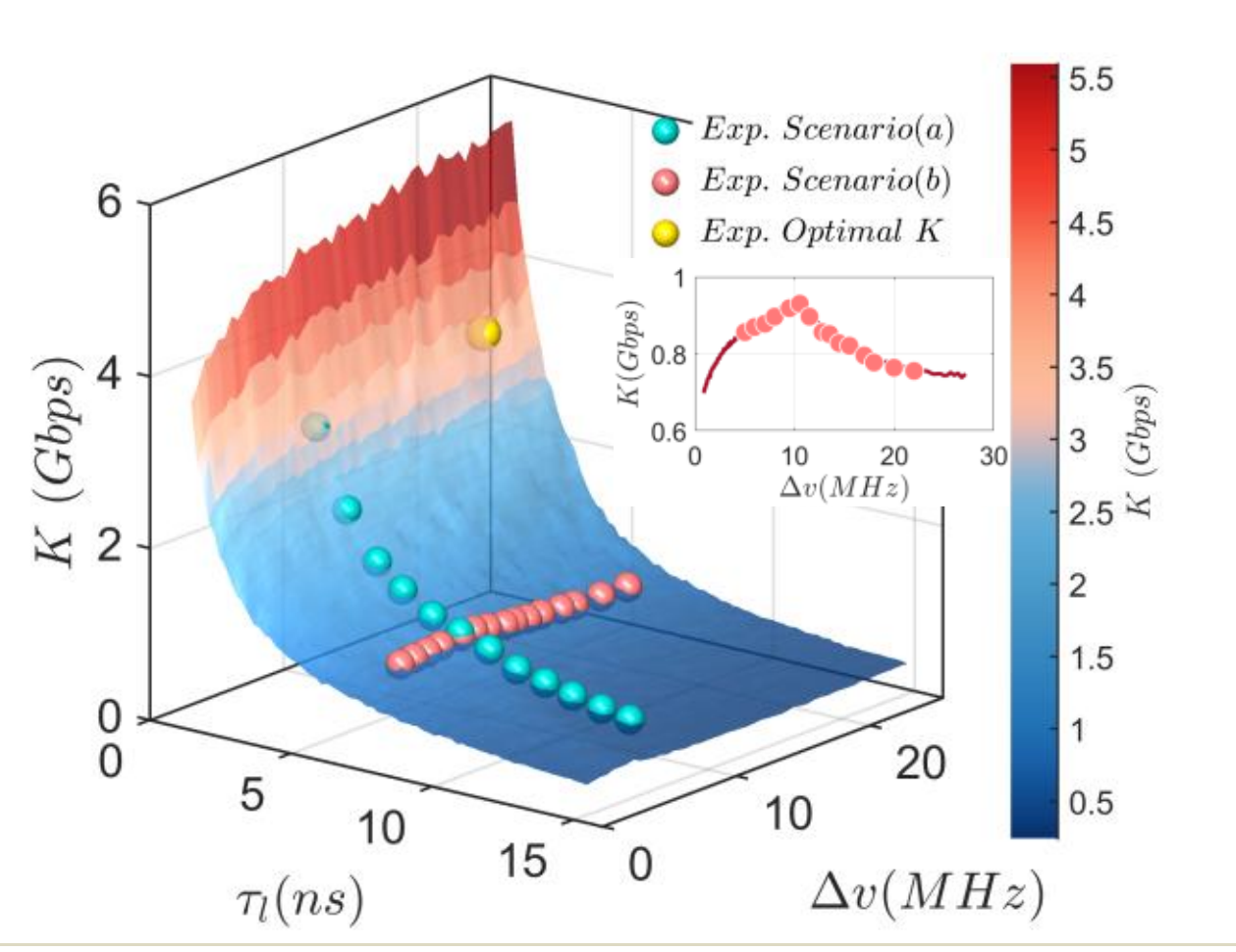

**Figure 7 Dependence of $K$ on $\tau_l$ and $\Delta v$. The color-coded surface represents the simulated results across the entire parameter space, where the color gradient indicates the magnitude of $K$. Experimental data for Scenario (a) and Scenario (b) are denoted by cyan and pink spheres, respectively.**

As indicated by the cyan spheres, $K$ exhibits a rapid initial decline with increasing $\tau_l$, followed by a more gradual decrease. This behavior stems from the fact that the reduction in $B_{ES}$ outweighs the corresponding increase in $H_{min}$ across the range. Consequently, the $K$ is overwhelmingly dominated by the $B_{ES}$. Notably, $K$ can not increase infinitely by simply reducing $\tau_l$. In the extreme limit of vanishing $\tau_l$, $H_{min}$ collapses to zero, which fundamentally forces $K$ to zero. A thorough investigation of the system dynamics in this ultra-short $\tau_l$ regime is reserved for the future work. As for scenario (b), indicated by the pink spheres in Fig.7, $K$ features a distinct peak, initially increasing slightly before decaying. This behavior is attributed to the fact that in the small-$\Delta v$ range, $B_{ES}$ remains nearly constant, thus, allowing $H_{min}$ dominates the overall trend of $K$.

Crucially, Fig. 7 highlights that the optimal $K$ is not achieved by merely maximizing $B_{ES}$ or $H_{min}$. Rather, achieving the optimal $K$ necessitates a delicate equilibrium between these two parameters. By calculating their product across the entire parameter space, we successfully identify the specific experimental configuration required for realistic performance optimization, which is marked by the bright yellow sphere in Fig.7.

The proposed predictive model successfully guides the physical configuration of the system to its optimum. Specifically, in this work, the maximum $K$ is experimentally achieved at $\Delta v = 22$ MHz and $\tau_l =$ 1.4 ns, yielding an entropy source bandwidth of $B_{ES}$ =255.4 MHz and min-entropy of $H_{min} =$ 6.43 bits per sample. After sampled at 500 MSa/s, the raw data is subjected to a 2048×1600 Toeplitz matrix to extract the final secure randomness. This optimal configuration delivers a final random number generation rate of 3.13 Gbps. As detailed in Fig,8, the post-processed data passes the standard statistic tests of NIST, confirming the security of the generated sequences.

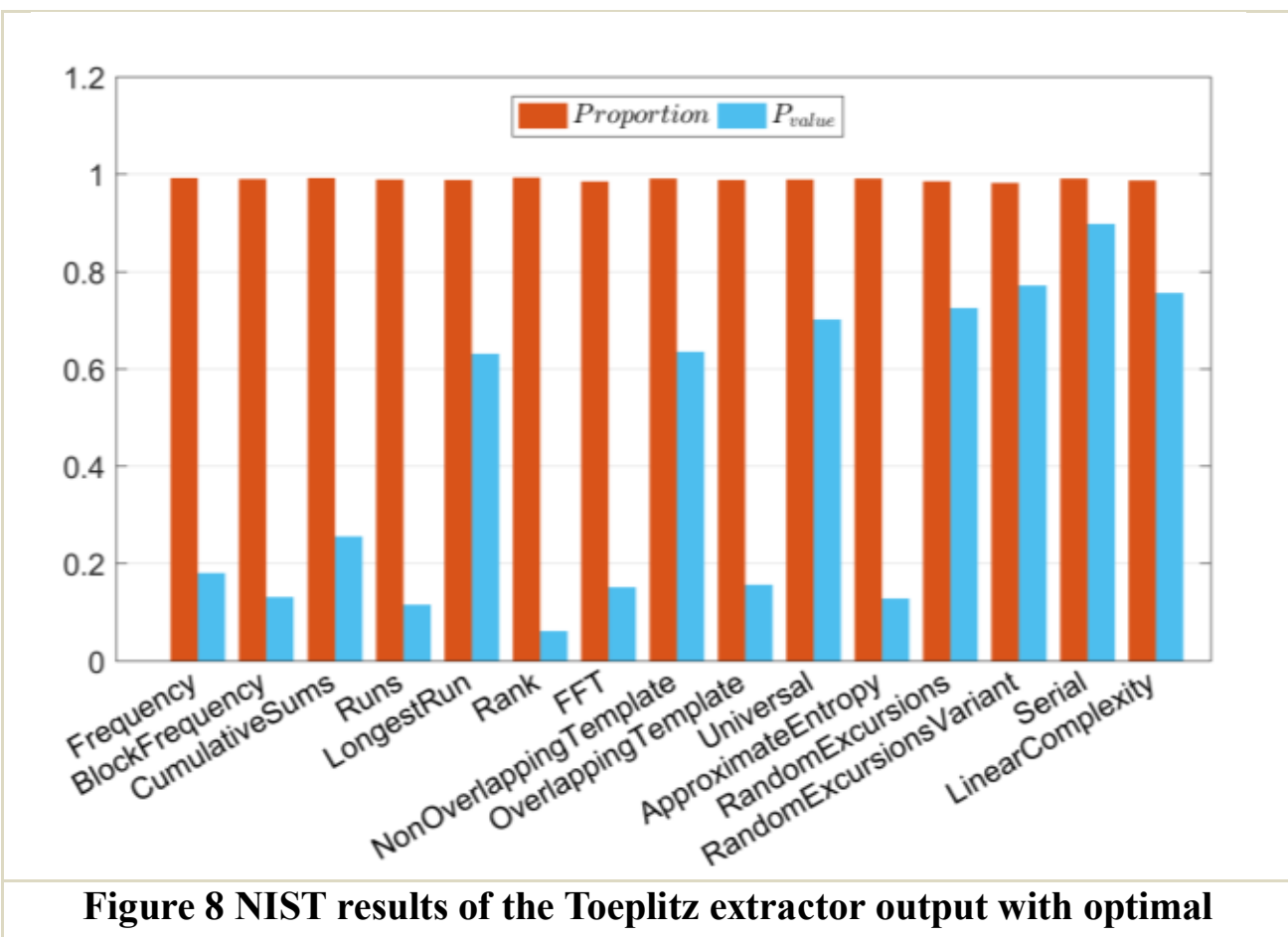

**Figure 8 NIST results of the Toeplitz extractor output with optimal random number generation rate**

# IV. Conclusion and Discussion

In this work, a comprehensive model based on a discrete time series of Wiener process is proposed to quantify the system performance of LPN-based QRNG. Based on the model, the comprehensive property of entropy source, including the power spectrum and probability distribution are estimated and thus the entropy source bandwidth and min-entropy under a given ADC are accurately calculated, which, as far as we know, is analyzed for the first time. The significant agreement between the simulated and experimental results validates the accuracy and reliability of the proposed model.

Furthermore, it is assumed in this work that the raw data acquired at a sampling rate of twice the entropy source bandwidth is i.i.d. For scenarios where the sampled data exhibit non-i.i.d. characteristics have been investigated in our parallel study, as detailed in Ref [41].

The proposed model offers an approach for the design of components in the LPN-based QRNG system, including laser linewidth, uMZI path delay difference, photodetector bandwidth, and ADC quantization accuracy and range, to achieve the comprehensive optimization based on the specific application requirements. Most significantly, this approach substantially reduces system development costs, particularly for photonic integrated QRNG research which generally requires iterative performance testing and validation.

# Funding

This work was supported by the National Cryptologic Science Fund of China (Grant No. 2025NCSF02053), the Quantum Science and Technology-National Science and Technology Major Project (Grant No. 2021ZD0300701), the National Natural Science Foundation of China (Grants Nos. 62171446, 62171418, 62201530, U24B20135, 62301517), the Industrial Prospect and Key Core Technology Projects of Jiangsu Provincial Key R & D Program (Grant No. BE2022071), the National Key Laboratory of Security Communication Foundation (2026, 6142103042603) the Stability Program of National Key Laboratory of Security Communication (WD 202501).

The authors are grateful to Professor Fangxiang Wang for his valuable suggestions during the development of this work, which improved the clarity of the discussion regarding the model's applicability.

# Disclosures

The authors declare that there are no conflicts of interest related to this article.